\documentclass[12pt]{iopart}
\usepackage{graphicx}
\usepackage{epstopdf}
\usepackage{amsfonts}
    
\begin{document}

   \title{G\"{o}del's incompleteness theorem and universal physics theories}

   \author{Uri Ben-Ya'acov}

   \address{School of Engineering, Kinneret Academic College on
   the Sea of Galilee, \\   D.N. Emek Ha'Yarden 15132, Israel}

   \ead{uriby@kinneret.ac.il}

\vskip 2.0cm

\begin{abstract}
 An ultimate universal theory -- a complete theory that accounts, via few and simple first principles, for all the phenomena already observed and that will ever be observed -- has been, and still is, the aspiration of most physicists and scientists. Yet, a basic principle that is embodied in the results of G\"{o}del's incompleteness theorems is that self-referencing leads to logical conflict or failure, as in the liar paradox or Russell's paradox. In physics theories self-referencing necessarily occurs when it is realized that the observer is also a participant in the experienced phenomena -- we, humans, are part of the universe while observing it. Therefore self-referencing, and consequently logical conflicts, are unavoidable, and any theory pretending to be universal is bound to be incomplete.
\end{abstract}



\vskip30pt

\noindent\textit{``A human is the universe's way of knowing the universe''}

\hskip250pt\emph{Anonymous}

\vskip30pt
\eject

\section{Introduction}\label{sec: intro}

Is it possible to encompass the full extent of the universe with a finite number of first principles and inference rules ?

The belief that it is possible to arrive at a complete theory that fully describes the whole of the physical world -- a theory that accounts, via few and simple first principles, for all the phenomena already observed and that will ever be observed -- has been, for many-many years and for most researches, a fundamental tenet of the scientific research.

Such a theory, sometimes known as ``theory of everything'' (TOE), is expected to be the ultimate theory of the universe. A. Einstein put it very clearly :

\begin{quote}
\textit{It is the grand object of all theory to make these irreducible elements as simple and as few in number as possible, without having to renounce the adequate representation of any empirical content whatever.} \cite{Einstein}
\end{quote}

By ``these irreducible elements'' Einstein referred to the fundamental first principles of the theory. It is very clear that in spite of all the hopes and aspirations, we are still very-very far from reaching such a goal. Quoting also S. W. Hawking :

\begin{quote}
\textit{.. How far can we go in our search for understanding and knowledge? Will we ever find a complete form of the laws of nature? (a complete form = a set of rules that in principle at least enable us to predict the future to an arbitrary accuracy, knowing the state of the universe at one time.)}

\textit{Up to now, most people have implicitly assumed that there is an ultimate theory that we will eventually discover. Indeed, I myself have suggested we might find it quite soon.} \cite{Hawking02}
\end{quote}

However, G\"{o}del's incompleteness theorem \cite{Godel,Hofstadter,NagelNewman,Raatikainen} implies that \textit{any formal structure, based on a finite number of first principles and inference rules, which is rich enough, cannot be at the same time both consistent and complete}. Completeness, in terms of physics theories, implies determinism -- if the necessary initial data are given then the theory allows us to predict the state of a physical system any time in the future.

Does G\"{o}del's theorem apply to physics ?

A common argument in favour of applying G\"{o}del's theorem to physics, is, more or less, that ``G\"{o}del's theorem applies to arithmetics which is the basis of mathematics, physics uses mathematics, therefore G\"{o}del's theorem applies to physics'' \cite{Jaki04,Jaki06}. However, the counter-argument points to the fact that there are mathematical theories to which G\"{o}del's theorem does not apply, \textit{e.g.} geometry \cite{Raatikainen}, and that this is the type of mathematics that physics uses, therefore we should not expect that G\"{o}del's theorem applies to physics \cite{Barrow,Feferman06,Robertson00}.

The purpose of this article is to present and put forward another argument, which I believe is the decisive one. G\"{o}del's theorem (whenever it applies) points to incompleteness in the sense that \textit{there will always be claims that may be formulated within this formal system but are undecidable} -- propositions that cannot be either proved or refuted. A close inspection of G\"{o}del's theorem demonstrates that this undecidability arises when the claims are \textit{self-referential}, with the system asking to define itself in its own terms, allowing paradoxical self-negation resulting in logical conflicts.

Self-referencing occurs in physics whenever the observer is also part of the observed system. In most physical systems this is not the case, therefore the counter-argument should apply. However, when it is the whole universe that is dealt with, then we, the observers, are also part of it, and G\"{o}del's theorem should apply.

The article discusses self-referencing in G\"{o}del's theorem, its relation with our involvement in the universe, its application to physics theories, and the eventual consequences -- the impossibility of a finite ``ultimate universal theory'' or ``theory of everything''.

\vskip30pt

\section{G\"{o}del's incompleteness theorem and physics theories}\label{sec: Godelphys}

Structurally, physics theories are expected to be complete formal (logical) systems, with a finite, consistent set of fundamental principles (axioms) and a set of deduction and inference rules which together produce predictions regarding natural phenomena; and completeness implies the expectation that the predictions of the theory encompass all the phenomena already observed and those that will ever be observed, within the realm covered by the theory.

Such theories may be likened to trees : The basis of the theory, the set of fundamental principles, is like the roots of the tree; the products and predictions of the theory are like the leaves, flowers and fruits of the tree; and the rules of inference and deduction are like the trunk and the branches which lead the sap from the roots to the leaves and flowers. In a tree the sap flows up by the force of capillarity; in a theory deductions are made from first principles to predictions by the power of logic.

A fundamental property of logic is embodied in G\"{o}del's incompleteness theorem \cite{Godel,Hofstadter,NagelNewman,Raatikainen} which implies that

\begin{quote}
\textit{In any consistent and rich enough formal structure, based on a finite number of first principles and inference rules, there will always be propositions that may be formulated within this formal system but are undecidable. Such a theory cannot be both consistent and complete.}
\end{quote}

G\"{o}del's theorem has been discussed heavily in the contexts of mathematics, logic, and artificial intelligence \cite{Raatikainen,Lucas,Penrose,Feferman09}. It's relation to physics seems to be much less discussed. Clearly, if G\"{o}del's theorem applies to a physics theory then this theory cannot be complete -- there will always be propositions that may be formulated within the theory but are undecidable because of self-referencing that leads to logical conflicts. Incompleteness in physics theories implies non-determinism -- the existence of processes whose outcome cannot be predicted by the theory.

As pointed out in the Introduction, a general argument raised for the applicability of G\"{o}del's theorem to physics is that since physics theories use mathematics and logic they cannot be complete \cite{Jaki04,Jaki06}. However, there are various mathematical theories ({\textit{e.g.}, geometry or reduced versions of arithmetics) \cite{Raatikainen} to which G\"{o}del's theorem does not apply -- they are too simple for that and are, therefore, complete. The counter-argument that is brought up then is that the mathematics used for the physics theories we know so far is relatively simple, not suffering from incompleteness, so there is no place to apply G\"{o}del's theorem to physics \cite{Barrow,Feferman06,Robertson00}.

These arguments miss the main aspect of G\"{o}del's theorem -- that undecidability, indicating a logical failure, appears there as a result of self-referencing. Indeed, several authors linked G\"{o}del's theorem with cases of undecidability in physics, especially quantum theory \cite{Komar,PeresZurek,Peres}. However, there may be various reasons for undecidability, not necessarily associated with self-referencing and logical conflicts. Yet, self-referencing is very relevant to physics theories, because we, humans, the observers, are also participants -- we are part of the universe while observing it. In the following we discuss self-referencing in relation to G\"{o}del's theorem, and then its relevance to physics theories.

\vskip30pt

\section{Self-referencing and G\"{o}delean propositions}\label{sec: SelfrefGodel}

A core principle of Aristotelian logic is \textit{the principle of excluded third}, expected to be obeyed by all formal (logical) systems \cite{Gottlieb,Horn}: \textit{Any claim or prediction that may be formulated within the theory is decidable -- can be either proved or refuted, no other option}. Yet, G\"{o}del's incompleteness theorem identifies the existence of a special kind of claims and propositions, which are well phrased in a given logical system, but contrary to expectations can neither be proven correct nor refuted as wrong. Such claims and propositions are referred to in the following as ``G\"{o}delean'', and with them the theory necessarily becomes incomplete.

Certainly, not every unanswerable proposition is G\"{o}delean. G\"{o}delean propositions are undecidable not because of lack of data but because of the way they are formulated within the formal system -- \textit{self-reference} allowing paradoxical self-negation leading to a conflict which the formal system, with all its axioms and first principles, cannot resolve:

In G\"{o}del's theorem itself, it is expressing both arithmetical statements and meta-arithmetical statements ({\it i.e.}, statements {\it about} arithmetics) in the same language which creates the possibility of self-reference, allowing paradoxical self-negation with inevitable logical conflict. More adaptable to physics, this basic principle underlying G\"{o}del's theorem is illustrated with some well-known so-called logical paradoxes \cite{Beall.etal}: the liar paradox, the barber paradox, Russell's paradox, \textit{etc.} (see Appendix). A close examination of G\"{o}del's theorem and these paradoxes highlights that G\"{o}delean propositions are self-referential, and appear when the system seeks to define itself in its own terms: Is the man a liar or speaking the truth? Does the barber shave himself or not? Is the set $M$ (the set of all sets that are not members of themselves in Russell's paradox) a member of itself or not? -- then the system turns out to be unable to provide an unambiguous answer, and this inability manifests as a logical failure.

A very impressive graphical visualization of the logical conflict arising due to self-reference is offered by the drawing ``Drawing Hands'' of the Dutch artist Escher\footnote{M.C. Escher. His works can be found on the official website http://www.mcescher.com/ and on a wide variety of other websites.} -- the logically impossible realization of two hands mutually drawing each other (\Fref{fig:hands}).
\begin{flushleft}
\begin{figure}[h]
  \begin{minipage}[b]{250pt}
  \begin{flushleft}
  \caption{``Drawing Hands''\\M.C. Escher (1948)}\label{fig:hands}
  \end{flushleft}
  \end{minipage}
  \hskip20pt
  \includegraphics[width=5cm]{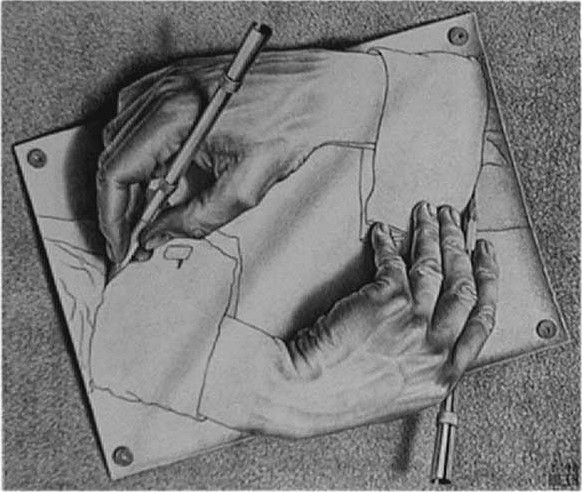}
\end{figure}
\end{flushleft}

Hence the conclusion that when such self-reference occurs within a formal system it necessarily leads to logical failure -- logical systems are unable to refer to themselves logically \cite{Bolander}. This is a fundamental trait of classical logic, implied by G\"{o}del's theorem. Thus, by means of G\"{o}del's theorem logic points to its own limitations -- the `third' cannot be excluded, and sufficiently large formal structures cannot be both consistent and complete.

\vskip30pt

\section{Self-referencing and physics theories}\label{sec: Selfrefphys}

By G\"{o}del's theorem, a theory needs to be sufficiently rich and large enough to allow self-referential, G\"{o}delean propositions. It depends on the domain encompassed by the theory. Classical theories, like geometry, Newtonian mechanics (excluding the question of relative/absolute space and time), Maxwell's electrodynamics, are free from self-referencing. Such theories are simply not large enough –- the observer is then no more than an outside witness, not involved at all, completely separated from the subject matter, and the theories may therefore be complete.

The case is essentially different with theories that aspire to encompass the whole universe, covering all the phenomena in nature. A universal theory, large enough to encompass the whole of the universe, must include the observers as part of its subject matter -- after all, we reside in the universe and are part of it. Once the inquirers are themselves part of the subject matter, self-referencing is unavoidable.

This was already recognized by Hawking some years ago,
\begin{quote}
\textit{We and our models are both part of the universe we are describing. Thus a physics theory is self-referencing, like in G\"{o}del’s theorem. One might therefore expect it to be either inconsistent or incomplete ..} \cite{Hawking02}
\end{quote}
and may be understood in several, inter-connected, ways :

In most observations the observer seems to be only a witness, and the physical process seems to be independent of the observer. But if the observer intervenes with the physical process, say, by determining the mode of observation (as in some quantum experiments, \textit{e.g.}, `wave or particle', or `one slit or two slits'), thus determining the outcome of the process, this is an act of self-reference.

Self-referencing requires two for tango -- and indeed, relativity theory (RT) emphasizes the fact that any observation is observer-dependent, depending on an arbitrarily chosen point-of-view (frame of reference); and quantum theory (QT) tells us that results of observations depend on an arbitrarily chosen mode of observation. These arbitrary choices are not controlled or determined by current physics theories (RT, QT) -- they (these arbitrary choices) are metaphysical statements about physical processes, like the meta-arithmetical G\"{o}delean statement in G\"{o}del's proof. Therefore, these arbitrary choices together with the processes they refer to, constitute self-referencing couples.

Self-referencing is directly associated with the evolution of the universe: As in the two-slit experiment, the mode of observation determines the fate of the physical process, thus directly affecting the evolution of the universe. Victor Hugo once said that \textit{Creation lives and evolves; the human is only a witness}. This was the common viewpoint up until the 20$^{\rm th}$ century. Now it becomes evident that we are not simply bystanders on a cosmic stage -- we are active participants in the evolution of the universe, the cosmos being made real in part by our own observations. This is the viewpoint put forward by people like E. Wigner and J. A. Wheeler, referring to observations of quantum phenomena. It is certainly reminiscent of the ancient Jewish tradition, that the human is participant in Creation.

Therefore, another feature of self-referencing in physics is that our observations contribute to the creation of physical reality. But then, the observer and his deciding faculty are also part of the universal reality, and should be included in a theory pretending to encompass the whole universe. Modern physics theories (RT, QT) are still not self-referential, because they do not include the observer. Such an inclusion should be an integral part of any futuristic, new-generation universal physics theory.

\vskip30pt

\section{Can G\"{o}del's theorem be circumvented ?}\label{sec: canGodel}

Since acts of observation may affect physical reality, these acts must be covered by the physics theory, which must be, therefore, self-referential. While current physics theories do not include the observers as part of their subject matter, futuristic theories, pretending to be ultimately universal or `theories of everything', must take the observers into account, and therefore will have to be self-referential. Inevitably G\"{o}del's incompleteness theorem will have to apply, rendering the physics theory incomplete and therefore non-deterministic.

We arrive, therefore, at the inevitable conclusion that there cannot be an ultimate, all-encompassing, universal physics theory which is founded upon few and simple, finite number of first principles.

The dissatisfaction that many people feel with this conclusion lead to looking for ways to bypass, or get around, this outcome of G\"{o}del's theorem \cite{Barrow,Feferman06}. Is it really possible, in some way, to remedy G\"{o}delean conflicts, especially those that appear in a universal theory?

The fact that current physics theories do not include the observers as part of their subject matter, and are therefore not self-referential, cannot be used as an escape route, because, as argued above, the universal physics theory will have to include the observers.

Another possibly considered escape route is the following:

The inability of a theory to decide regarding certain propositions indicates that fundamental principles are missing. Physics theories, besides being formal logical structures, have to stand the test of physical reality. G\"{o}delean undecidability is a feature of the theory, as distinct from the physical reality: When a phenomenon is observed, or an experiment is performed, there could be some questions that cannot be answered by the theory, while they are answered by the physical reality via the results of the experiment. Therefore, a G\"{o}delean conflict may be, ostensibly, solved by adding new axioms -- additional, new fundamental principles that are in concordance with phenomena and the results of experiments -- that sort out the conflict. This was the philosophy behind the (unsuccessful) hidden-variables approach. But then, as already observed by G\"{o}del, the result is another, larger, new theory, in which new G\"{o}delean propositions ensue. Attempting to solve G\"{o}delean conflicts in this way will be a never-ending series of adding new `first principles'. G\"{o}delean conflicts are inescapable! \cite{Raatikainen}

Thus another recognition by Hawking:
\begin{quote}
\textit{Maybe it is not possible to formulate the theory of the universe in a finite number of statements. This is very reminiscent of G\"{o}del's theorem ..} \cite{Hawking02}
\end{quote}

The many attempts to remedy the self-referencing paradoxes in the field of logic involve two main ingredients \cite{Beall.etal,Bolander}:
\begin{itemize}
  \item {Extending the classical binary range of truth values of `1' and `0' or `true' and `false', by adding new truth values like `undefined' (neither `true' nor `false'), `both true and false', or even a whole range of numerical truth values between 0 and 1.}
\end{itemize}
The extended range of truth values may correspond, physically, to, say, viewing the electron as both particle and wave, or to the possible spectrum of results of a quantum experiment.
\begin{itemize}
  \item {Regarding the realm upon which the theory operates as an hierarchical structure, a set of ordered levels, so that referencing may occur only from a higher level to a lower one.}
\end{itemize}
Self-reference paradoxes start with two entities, say $a$ and $b$, with a unidirectional hierarchical relation of {\it superiority} $b \searrow a$ between them (\textit{e.g.}, $b$ declares attribute $a$; $b$ shaves/doesn't shave $a$; $a$ is/isn't a member of set $b$; $b$ draws figure $a$; {\it etc.}). If $a$ and $b$ belong to two different spaces, say $a \in A$ and $b \in B$, so that each entity in $B$ is superior to each entity in $A$ then there is no paradox. But if $a$ and $b$ belong to the same space then the opposite relation $a \searrow b$ is also possible, and self-reference paradoxes become possible.

Therefore, self-reference paradoxes can be remedied only in a hierarchical structure. This is reminiscent of the solution to self-referencing in Russell's paradox within set theory: Construction of an infinite hierarchy of sets, with sets at a certain level being the members of higher-level sets \cite{Bolander}. Paradoxes are indications of misconceptions, and self-reference paradoxes appear when spaces or levels that should be separated are instead squeezed into just one space or level.

Self-referencing necessarily involves human consciousness\footnotemark : Inquiries, whether regarding a particular phenomenon or the entire universe, and decisions regarding the mode of observation, all take place within the realm of our consciousness. We -- humans, the observers -- we are all, with our consciousness, part of the universe. We observe and inquire the universe and our part in it \textit{from} and \textit{within} our consciousness. Therefore, questions regarding the universe are asked from within it, and at least some of the questions that we may ask about the universe and natural phenomena are self-referential.

\footnotetext{\textit{Mind} and \textit{consciousness} are meant here, in a broad sense, as \textit{the domain where mental processes take place and we interpret our experiences, whether internal or external, and find meaning and significance for them}.}

This suggests that our consciousness must be seen as hierarchically structured, so that consciousness refers to itself from a high level to a lower one: Entities that appear separate at a certain level seem to be, from a higher level of consciousness, part of a single complex system (as the {\it whole} and its {\it parts}). This, at least potentially, is an infinite hierarchy, which, as we know from set theory, grows exponentially.

\vskip30pt

\section{Can physics be complete ?}\label{sec: physcomp}

The centuries-old vision for physics is the expectation to arrive at a final theory, based on few first principles, that can describe and explain the whole physical world and all that is attached to it, past present and future.

However, is the physics that we know today (or that we may know at any given time) all that there is to know? Can we ever be sure that essentially new phenomena will never be observed any more?

We certainly can’t. History definitely tells the opposite. Yet the vision is very strong, and very domineering upon the scientific community

Can we at least predict, speculate into the future, regarding the next phase of physics? Can we get a glimpse into what is currently beyond the horizon of physics? Can we get hints from what is already known?

\vskip15pt

Any theory that aspires to encompass the whole universe, with all the phenomena in nature, is the result of the universe, via human consciousness, inquiring itself. We (humans) are part of the universe, and physics (or science, more generally) is our (human) interpretation of what we observe in the universe. Acts of observation should therefore be considered active part of physical processes, hence part of the subject matter of physics theories.

We are therefore both observers and observed. Moreover, we may observe, being aware, our own observations. Since in an act of observation the observer refers to the observed phenomenon, thus observation is referencing in physics, and self-observations are self-referencing in physics.

Self-referencing is therefore unavoidable in physics. Indeed, current physics theories do not include the observers and their observations as part of their subject matter, and are therefore not self-referential; but as argued above, the universal physics theory will have to take the observers and observations into account, and therefore will have to be self-referential.

The foregoing discussion thus indicates that regarding the universe as one big whole realm then the corresponding universal theory must allow G\"{o}delean self-referencing propositions. Consequently, the theory is necessarily incomplete -- there will be issues of physical reality about which the theory cannot make predictions.

When a new phenomenon is observed that the current theory cannot explain, we seek for new first principles that cover these new observations. But the addition of new first principles cannot solve the G\"{o}delean conflict, because the new theory will still be the universe inquiring itself, and self-referencing unavoidable. Thus, more and more new first principles will have to be added, {\it ad infinitum}.

That some new observations require new first principles has always been the drive for scientific advancement. In this way the scientific research produces more and more insights, understandings and knowledge, within larger and larger theories. Still, many people wish to arrive, hopefully in their life-time, to a final finite theory that fully describes, with few and simple first principles, the whole of the physical world, hoping and believing in the vision put so clearly by Einstein in the quotation brought in the Introduction. But G\"{o}del's theorem indicates that this is a never-ending process.

Therefore, it is impossible to encompass with a finite number of first principles and inference rules the full extent of the universe. For any final set of first principles there are, or will be found, some issues about which the theory is unable to make predictions.

\vskip30pt

\section{Concluding remarks}\label{sec: concrem}

The expectation for a finite TOE rests on the ancient belief that the universe is fixed and eternal. But as we have learnt from modern physics, which emphasizes the rôle of the observer and where at least some predictions are observer-dependent, our inquiries and observations affect the evolution of the universe. Therefore the universe is not fixed, but constantly evolving, with the possibility of new phenomena appearing. The ``cake of knowledge and understanding'' is not fixed, unchanging, but rather ever-growing, with new insights, new knowledge, new information, that are not derivable from old ones. Thus there will always be more to be known, to be revealed, to discover; not less.

In conclusion, there will always be some questions seeking the understanding of the universe as a whole which will be answerable only partially. Each new answer will enlarge our understanding, but at the same time will make us realize how little we actually know and how much more there is to be known and understood. The more we know then there is, and will be, even more to be known and reveal. The prospects to discover become larger, not smaller.

Thus the recognition that ``\textit{The more we know, the more we know that we know less .. and it will never stop !}'' is unavoidable. The search for natural laws is, and will always be, open-ended.

The impossibility of a finite ``ultimate universal theory'' or ``theory of everything'' (with the understanding that a theory cannot be based on an infinite number of first principles), and the fact that it is organically inherent in the nature of the scientific process, as is asserted by G\"{o}del's theorem, is not realized by many, and when it is it seems, unfortunately, to cause much disappointment \cite{Barrow,Feferman06}. However, if a finite ultimate theory that fully describes, with few and simple first principles, the whole of the physical world, were possible, then what new first principles would be left for the coming generations to reveal and discover? What prospects would they have then? Thus we end with one more quotation of Hawking:
\begin{quote}
\textit{Some people will be very disappointed if there is not an ultimate theory that can be formulated as a finite number of principles. I used to belong to that camp, but I have changed my mind. I'm now glad that our search for understanding will never come to an end, and that we will always have the challenge of new discovery. Without it, we would stagnate. G\"{o}del’s theorem ensured there would always be a job .. for physicists.} \cite{Hawking02}
\end{quote}
I agree.

\vskip30pt

\appendix

\section{}

Here is a sample of the most well-known logical paradoxes caused by self-reference:

\subsection{The liar paradox}

The most basic and best-known of the logical paradoxes, telling of someone who declares of himself ``I am a liar''. Formally, the paradox consists in the declaration ``this sentence is false'', so that if only two truth values are acceptable (true = 1, false = 0) the sentence is well formulated in the language of logic but has no logical sense. This is a paradox of self-testimony.

A basic formal extension of the liar's paradox is structured as follows:

	``A: Sentence B is true'', ``B: Sentence A is false.''

Here every sentence is, separately, logically well constructed, but their combination leads to a conflict. It is possible, of course, to extend this presentation to any number of sentences, so that the logical failure is revealed only in the overall array.

\subsection{The barber paradox}

The story goes of a small French village, in which there is only one barber, and he shaves every day all the men that do not shave themselves.
Then the question is, since the barber also needs to shave, ``Who shaves the barber?''

The sentence ``every morning Mr. Jean-Pierre enters the barber's shop and gets a shave'' is very well defined within the context of the barber's story. It leads to a conflict only if we are told that Mr. jean-Pierre is the barber himself, {\it i.e.}, when it becomes a self-referencing proposition.

\subsection{Russell's paradox}

The British philosopher Bertrand Russell challenged the question ``Can a set be a member of itself?'' by defining a set $M$ as ``the set of all the sets that do not contain themselves as their own members''. If $M$ is a member of itself, then it cannot be, and vice versa. The conflict shows that such a set is impossible, and thus defies the possibility that a set may contain itself as a member of itself.

\vskip30pt

\rule{10cm}{1pt}


  \end{document}